\documentclass[11pt]{article}
\usepackage{amsmath}
\usepackage{amsfonts}
\usepackage[utf8]{inputenc}
\usepackage[T1]{fontenc}
\usepackage{graphicx}

\newcommand{\EE}{\mathbb{E}}
\newcommand{\PP}{\mathbb{P}}

\newcommand{\VV}{\mathbb{V}}

\newcommand{\VaR}{\operatorname{VaR}}
\newcommand{\Vas}{\operatorname{Vas}}

\begin{document}

\title{Model Risk in Credit Portfolio Models}
\author{Christian Meyer\footnote{DZ BANK AG, Platz der Republik, D-60265 Frankfurt. The opinions or recommendations expressed in this article are those of the author and are not representative of DZ BANK AG.} \footnote{E-Mail: {\tt Christian.Meyer@dzbank.de}}}
\date{\today}
\maketitle

\begin{abstract}
Model risk in credit portfolio models is a serious issue for banks but has so far not been tackled comprehensively. We will demonstrate how to deal with uncertainty in all model parameters in an all-embracing, yet easy-to-implement way.
\end{abstract}

\section{Introduction}

At present, model risk is ranking high on the financial industry's agenda. Papers are being written, conferences are being organised, and regulators around the world are trying to establish rules for model risk management. The US regulatory authorities have been leading in this respect, with the OCC and Fed already having proclaimed in 2011 (cf. \cite{OCC2011}, \cite{Fed2011}, also \cite{OCC2021} and now \cite{Fed2026}) that ``model risk should be managed like other types of risk''.

Among the models employed by banks, those used for calculation of capital requirements are of particular importance, both for the institution and for the industry as a whole. And furthermore, since for most banks credit risk is the major risk, the spotlight will be on model risk connected with credit portfolio models. However, literature on model risk in credit portfolio models is scarce (cf. \cite{RMV}, Part IV, for an overview).

How should one manage model risk of a credit portfolio model? The primacy of management over measurement is certainly sensible, but at least some quantification will be required to make management possible. Consider the OCC and Fed again: ``With an understanding of the source and magnitude of model risk in place, the next step is to manage it properly.'' In this respect, a holistic view of the model based on simple tools is more useful than elaborated machinery (coming with its own model risk) capturing only certain aspects.

In a nutshell, credit portfolio models usually take trade-level parameters, such as probability of default (PD) and loss given default (LGD), and then construct scenarios under the assumption of some concept of dependence between defaults, requiring additional parameters (think of ``correlations'').

Usually, all the parameters are given as point values (a PD of 1\%, an LGD of 50\%, a correlation of 10\%), and the main source of model risk (leaving aside, for the moment, the general assumption that a credit portfolio model can be constructed that way) is the uncertainty behind these parameters. This uncertainty can be of the epistemic type (How can they be observed or inferred? From what data?) or of the aleatory type (Even if the estimation process makes sense, what is the effect of the scarcity of data?). We propose, following Briggs \cite{Briggs}, that both types of uncertainty can be addressed by \emph{treating the parameters as additional risk factors}. That is, the point values are replaced by distributions, and the parameters are ``integrated out of the model''. We will focus on parameter distributions that appreciate the given point values by setting them as expectation values -- an assumption that might be challenged in a second step, of course.

Note that this procedure does not \emph{eliminate} the model risk but makes it \emph{measurable} and through this \emph{manageable}. For example, it becomes possible to compare the different model parameters and rank them according to their impact on model risk. This procedure can be enormously helpful for model validation. In particular, it goes a lot further than mere sensitivity analysis in that it addresses the problem of \emph{how far} to stress the parameters.

We will propose techniques for dealing with the parameters PD, LGD and correlation. We will explain these techniques using a well-known toy model. However, the methodology will be easily extendable to almost all credit portfolio models used in practice (cf the concluding section).

Throughout the paper we will denote the standard normal density by $\varphi(\cdot)$, the standard normal cumulative distribution function by $\Phi(\cdot)$, and the cumulative distribution function of the two-dimensional standard normal distribution with correlation coefficient $\varrho$ by $\Phi_2(\cdot,\cdot,\varrho)$. We will use the symbols $\PP$, $\EE$ and $\VV$ for probabilities, expectation values and variances.


\section{The toy model}

The credit portfolio model to be presented in this section is a standard toy model that serves well to illustrate all kinds of effects in credit risk modelling. We will comment on generalisation of results in the concluding section.

A single position is assigned an abstract asset value:
\begin{equation}
A = \sqrt{\varrho}\cdot X + \sqrt{1-\varrho}\cdot \epsilon
\end{equation}
Here, $X$ and $\epsilon$ are i.i.d. standard normal. $X$ is the systematic risk factor (shared with other positions, and possibly one of a set of correlated systematic factors), and $\epsilon$ is the idiosyncratic risk factor specific to the position. The parameter $0\leq\varrho\leq 1$ is often referred to as the asset correlation. By construction, the asset value $A$ is standard normally distributed as well. Default is assumed to occur if a default barrier $C$ is hit, which is calibrated to the probability of default $p$:
\begin{equation}
A \leq C, \qquad C := \Phi^{-1}(p)
\end{equation}
In such a model, defaults of different positions are independent when conditioned on realisations of the systematic risk factors. The probability of default conditional on $X = x$ reads:
\begin{equation}
\PP(A \leq C \,|\, X = x) = \Phi\left(\frac{\Phi^{-1}(p)-\sqrt{\varrho}\cdot x}{\sqrt{1-\varrho}}\right)
\end{equation}
A random variable
\[
L = L(X) := \Phi\left(\frac{\Phi^{-1}(p)-\sqrt{\varrho}\cdot X}{\sqrt{1-\varrho}}\right)
\]
with standard normally distributed $X$ is said to be Vasicek-distributed with parameters $p$ and $\varrho$,
\[
L\sim\Vas(p,\varrho),
\]
that is
\[
\Phi^{-1}(L) \sim N\left(\frac{\Phi^{-1}(p)}{\sqrt{1-\varrho}},\frac{\varrho}{1-\varrho}\right)
\]
with
\[
\EE(L) = p, \qquad \VV(L) = \Phi_2\left(\Phi^{-1}(p),\Phi^{-1}(p),\varrho\right) - p^2.
\]
Now, under additional assumptions (there is only one systematic risk factor; exchangeability; the portfolio is homogeneous with respect to $p$, very large and granular; for details cf. \cite{Gordy}) the variable $L$ can be interpreted as \emph{portfolio} loss quota. Assuming a loss given default quota $q$, we get a toy credit portfolio model $q\cdot L$ (albeit a quite serious one, since it is the basis of the Basel II IRBA formula, cf. \cite{Basel}) with three parameters: $q$, $p$, $\varrho$, all in the range $[0,1]$. The typical risk measure will be a Value-at-Risk, i.e., a one-sided $\alpha$-quantile. In the case of the toy model, the Value-at-Risk reads:
\begin{equation}
\VaR_{\alpha} = q\cdot \Phi\left( \frac{\Phi^{-1}(p)}{\sqrt{1-\varrho}} + \sqrt{\frac{\varrho}{1-\varrho}} \cdot \Phi^{-1}(\alpha)\right)
\end{equation}
We will now consider uncertainty in each of the parameters $q$, $p$ and $\varrho$. We will illustrate the effect on the toy model using a toy example of $p=0.01$ (consider a retail portfolio), $\varrho=0.1$ (correlations in that range are quite common), $\alpha=0.999$ (as in the Basel formula), and $q=0.5$ (again, LGD quotas in that range are common). Plugging these values into the Value-at-Risk formula, we get a value of $0.0387$.

\section{Uncertainty in the LGD}

Loss given default (LGD) quotas are usually given as expectation values. However, realizations of loss quotas around that value tend to be quite rare: either there is (almost) full recovery, or (almost) all is lost. It is therefore quite reasonable to treat default as a collection of different possible states, in analogy to rating migration. Sen \cite{Sen} and Krekel \cite{Krekel} have explored this in detail.

We will concentrate on the simplest, yet most extreme case and assume that actually there are only two possibilities: total recovery or total loss. In our setup, the first case is indistinguishable from non-default. In order to preserve the expected loss, the probability for the second case has to be set to $p\cdot q$, where $p$ is the probability of default and $q$ is the expected LGD quota. Note that if $q=0$, the discussion about model risk should take place on a different level.

Technically, we have transferred the factor $q$ from loss to PD. In the toy model, the Value-at-Risk therefore now reads:
\begin{equation}
\VaR_{\alpha} = \Phi\left( \frac{\Phi^{-1}(p\cdot q)}{\sqrt{1-\varrho}} + \sqrt{\frac{\varrho}{1-\varrho}} \cdot \Phi^{-1}(\alpha)\right)
\end{equation}
In the toy example ($p=0.01$, $\varrho=0.1$, $\alpha=0.999$, $q=0.5$), the Value-at-Risk rises from $0.0387$ to $0.0460$.

\section{Uncertainty in the PD}

Probabilities of default (PDs) usually have to be calibrated based on small samples. Sample size $N$ will be the main driver of uncertainty. One way to express the uncertainty behind a probability of default $p$ is via a $B(N,p)$ binomial distribution, scaled by $N$:
\[
N\cdot P \sim B(N,p)
\]
However, in order to integrate such a distribution into a credit portfolio model we would have to rely on simulation. Instead, we can look for alternative distributions on the interval $[0,1]$ that can be fitted to expectation value $p$ and variance $p\cdot(1-p)/N$. The beta distribution comes to mind for obvious (i.e., ''Bayesian'') reasons but we will assume a Vasicek distribution $PD~\sim \Vas(p,\beta)$ instead (in practice, the possible shapes those two types of distribution can take are very similar).

That is, we will assume that
\[
\Phi^{-1}(PD) \sim N\left(\frac{\Phi^{-1}(p)}{\sqrt{1-\beta}},\frac{\beta}{1-\beta}\right)
\]
and solve for $\beta$ by comparing the variances:
\[
\Phi_2\left(\Phi^{-1}(p),\Phi^{-1}(p);\beta\right) = \frac{p\cdot(1-p)}{N} + p^2
\]
What is the advantage? If the PD is embedded in another Vasicek-distributed variable $L~\sim \Vas(PD,\varrho)$,
\[
\Phi^{-1}(L) \sim N\left(\frac{\Phi^{-1}(PD)}{\sqrt{1-\varrho}},\frac{\varrho}{1-\varrho}\right)
\]
then (assuming independence) we find:
\[
\Phi^{-1}(L) \sim \Vas(p,\varrho + \beta - \varrho\cdot\beta)
\]
Therefore, in order to include the effect of PD uncertainty (and uncertainty in migration probabilities as well, if required) in the credit portfolio model, we can just tweak the correlation.

In the toy model, the Value-at-Risk then reads:
\begin{equation}
\VaR_{\alpha} = q\cdot \Phi\left( \frac{\Phi^{-1}(p)}{\sqrt{1-\varrho-\beta+\varrho\cdot\beta}} + \sqrt{\frac{\varrho+\beta-\varrho\cdot\beta}{1-\varrho-\beta+\varrho\cdot\beta}} \cdot \Phi^{-1}(\alpha)\right)
\end{equation}
To illustrate, Table \ref{table_Basel} assumes $N=1000$ (corresponding to, say, sample size 100 over a 10 year period -- I recommend using total sample size, but this is up to debate). For PDs from one basis point to 4096 basis points it presents $\beta$, the adapted correlation $\varrho + \beta - \varrho\cdot\beta$ based on $\varrho=0.12$, and the Basel correlation $0.12\cdot f(PD) + 0.24\cdot(1-f(PD))$ with
\[
f(PD) = \frac{1-\exp(-50\cdot PD)}{1-\exp(-50)}.
\]
We note that, corresponding to intuition, the effect of uncertainty is much larger for small probabilities of default. The Basel formula stresses correlations for small PDs as well but the form of the correcting function seems quite different.

\begin{table}
\begin{center}
\[
\begin{array}{rrrr}
\hline
p		& \beta	& \text{adapted} & \text{Basel}\\
\hline
0.0001	& 0.177	& 0.276	& 0.239\\
0.0002	& 0.140	& 0.243	& 0.239\\
0.0004	& 0.104	& 0.212	& 0.238\\
0.0008	& 0.073	& 0.184	& 0.235\\
0.0016	& 0.048	& 0.162	& 0.231\\
0.0032	& 0.030	& 0.146	& 0.222\\
0.0064	& 0.018	& 0.136	& 0.207\\
0.0128	& 0.011	& 0.130	& 0.183\\
0.0256	& 0.006	& 0.125	& 0.153\\
0.0512	& 0.004	& 0.124	& 0.129\\
0.1024	& 0.002	& 0.122	& 0.121\\
0.2048	& 0.002	& 0.122	& 0.120\\
0.4096	& 0.001	& 0.121	& 0.120\\
\hline
\end{array}
\]
\end{center}
\caption{Effects of PD uncertainty versus Basel correlation}
\label{table_Basel}
\end{table}

In the toy example ($p=0.01$, $\varrho=0.1$, $\alpha=0.999$, $q=0.5$), assuming $N=1000$ and therefore $\beta\approx 0.015$, the Value-at-Risk rises from $0.0387$ to $0.0431$.



\section{Uncertainty in the correlation}

Again, we assume that we have found a way to express the uncertainty in the correlation via some empirical distribution, e.g., as the outcome of a bootstrap exercise. Since the correlation will serve as variance of the systematic risk factor, we will be interested in distributions that produce something tractable when multiplied to a normal distribution.

If, for example, a random variable $R$ follows an inverse-chi-squared distribution with parameter $\nu$, another random variable $X$ is standard normal, and $R$ and $X$ are independent then $\sqrt{\nu\cdot R}\cdot X$ follows a $t$-distribution with $\nu$ degrees of freedom (cf. \cite{Schon} for an early mention of this idea in the context of credit risk modelling).

Since $\EE(R) = (\nu - 2)^{-1}$ and
\[
\VV(R) = \frac{2}{(\nu-2)^2 (\nu-4)}
\]
we find:
\begin{equation}
\nu = 4 + 2\cdot\frac{\EE(R)^2}{\VV(R)} = 4 + 2\cdot\left(\frac{\sigma(R)}{\EE(R)}\right)^{-2}
\end{equation}
Applying this to the toy model, we can address uncertainty in the correlation $\varrho$ by replacing the systematic factor $X$ by
\[
\sqrt{\frac{\nu - 2}{\nu}}\cdot X_{\nu},
\]
where $X_{\nu}$ is $t$-distributed with $\nu$ degrees of freedom, and $\nu$ is calibrated using the relative standard deviation of the empirical distribution. The factor $\sqrt{(\nu-2)/\nu}$ is optional and only serves to scale the variance to one. Note that the abstract asset value is no longer normally distributed (it is the weighted sum of a $t$-distribution and a normal distribution), and implicitly the PD changes (but usually not much, cf. the toy example below).

In the toy model, the Value-at-Risk then reads:
\begin{equation}
\VaR_{\alpha} = q\cdot \Phi\left( \frac{\Phi^{-1}(p)}{\sqrt{1-\varrho}} + \sqrt{\frac{\varrho}{1-\varrho}}\cdot \sqrt{\frac{\nu-2}{\nu}} \cdot T_{\nu}^{-1}(\alpha)\right)
\end{equation}
Here, $T_{\nu}$ denotes the cumulative distribution function of the $t$-distribution with $\nu$ degrees of freedom.

In practice, relative standard deviations of 0.25 are not uncommon for correlation distributions, resulting in $\nu = 36$ (it should be noted that in the context of IRC models, cf. \cite{EGIM} par. 153, the European Central Bank also requires test calculations using $\nu = 8$). In the toy example ($p=0.01$, $\varrho=0.1$, $\alpha=0.999$, $q=0.5$), the Value-at-Risk rises from $0.0387$ to $0.0425$ in this case. The PD increases slightly, from $0.01$ to $0.01001$. In real-life situations, the effect on the Value-at-Risk ist usually smaller due to idiosyncratic effects. In analogy to PDs, smaller correlations often lead to larger relative standard deviations, to smaller degree of freedom, and to larger uncertainty effects.


\section{Aggregation of effects}
\label{sec_aggregation}

\begin{figure}
\begin{center}
\includegraphics[width=12cm]{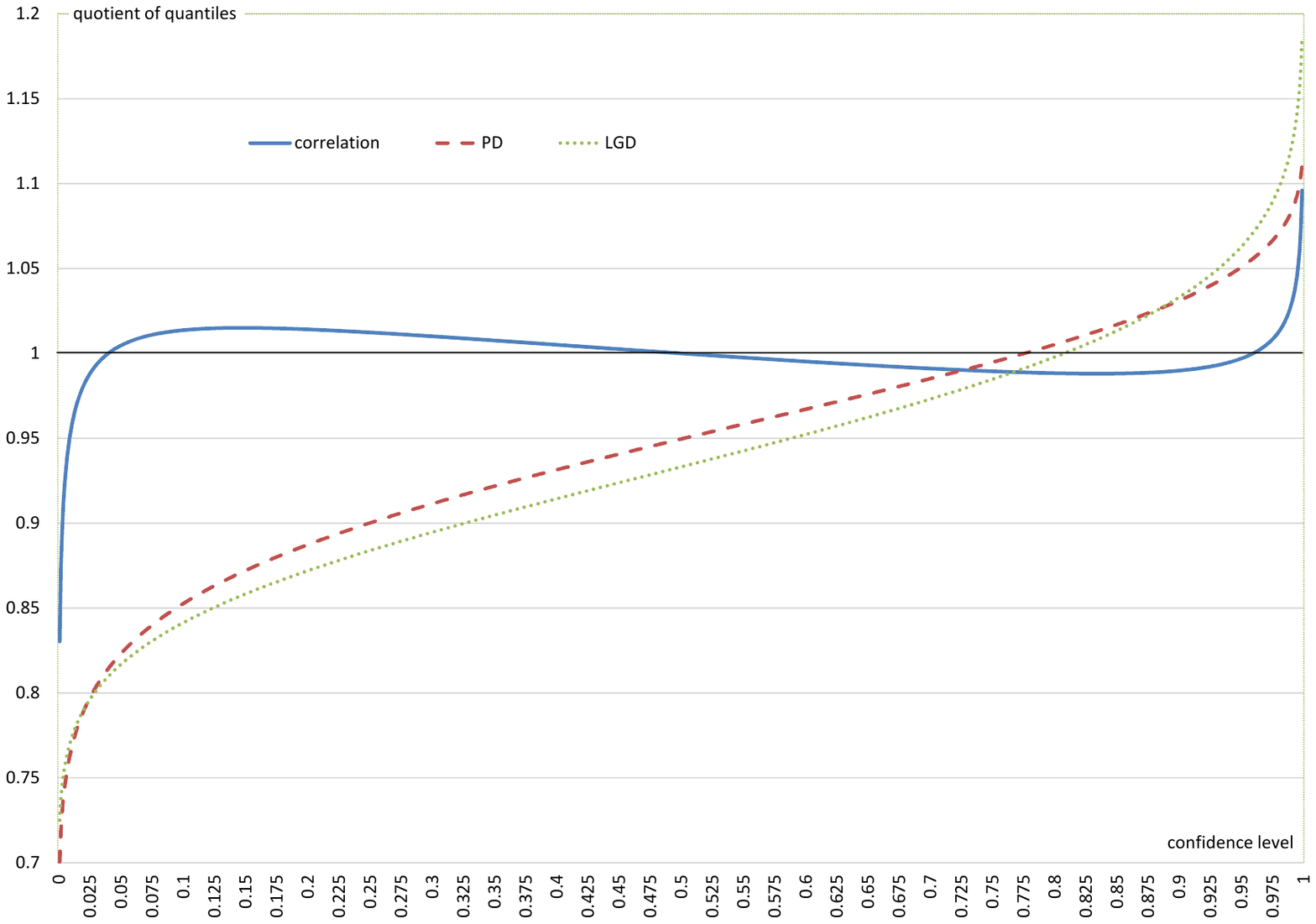}
\end{center}
\caption{Quotients of quantiles in the toy example}
\label{figure_QQ}
\end{figure}

So far we have been treating the parameters $q$, $p$ and $\varrho$ individually. To illustrate further, Figure \ref{figure_QQ} displays quotients of quantiles (quantile of adapted distribution divided by quantile of original distribution, for confidence levels between $0.001$ and $0.999$) in the toy example. Treatment of model risk for $\varrho$ means changing the distribution of the underlying risk factor, and has a different effect on the distribution than treatment of model risk for $q$ and $p$, which means adaptation of parameters.

Now, aggregation of model risk treatment for several parameters in the toy model is straightforward (just adapt the Value-at-Risk formulas iteratively). Table \ref{table_agg} displays the Value-at-Risk figures in the toy example for all combinations of parameters.

\begin{table}
\begin{center}
\begin{tabular}{lrclr}
\cline{1-2}\cline{4-5}
base case & 0.0387 & & $q$, $p$, $\varrho$ & 0.0574\\
$q$ & 0.0460 & & $p$, $\varrho$ & 0.0474\\
$p$ & 0.0431 & & $q$, $\varrho$ & 0.0510\\
$\varrho$ & 0.0425 & & $q$, $p$ & 0.0515\\
\cline{1-2}\cline{4-5}
\end{tabular}
\caption{Value-at-Risk figures for all combinations of parameters}
\label{table_agg}
\end{center}
\end{table}

\begin{figure}
\begin{center}
\includegraphics[width=12cm]{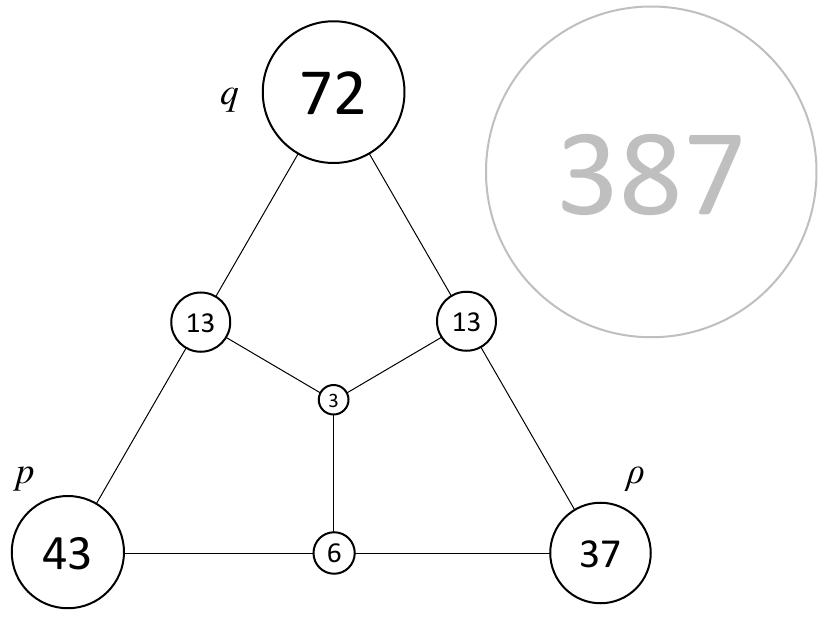}
\end{center}
\caption{Visualization of aggregated effects}
\label{figure_agg}
\end{figure}

At this point, it is worth taking some time to think about how to present results like these. For example, what is not immediately apparent from the table is that, in every single case, the effects are more than additive (in other words, there is no diversification of model risks in this setting). As an alternative in visualization (probably not the optimal one -- for example, it would not work if there \emph{was} diversification of model risks), I propose the diagram in Figure \ref{figure_agg}. Numbers are given in basis points, circle areas correspond to numbers, and numbers add up to total effect.

For example, the combined effect of uncertainty in PD and LGD is $43 + 37 + 6$ basis points, the $6$ being the more-than-additive effect. The total effect is 187 basis points, leading to an adjusted Value-at-Risk of 574 basis points.


\section{Conclusion}
\label{sec_conclusion}

We have proposed the following, wonderfully counter-intuitive procedure for dealing with model risk in the parameters of a credit portfolio model:
\begin{itemize}
\item If you want to address model risk in the LGDs, tweak the PDs (or, more generally, treat default as migration).
\item If you want to address model risk in the PDs, tweak the correlations.
\item If you want to address model risk in the correlations, change the distributional assumption of the systematic risk factors.
\end{itemize}
That leaves us with the question of model risk in the distributional assumptions of the risk factors. At this point, tweaking the model will not help. Instead, a benchmark model might step in, e.g., one based on networks or other contagion mechanisms. Or this might be a place for Artificial Intelligence methods.

We have illustrated the methodology using a toy model. However, it can easily be extended to credit portfolio models of CreditMetrics type (the de facto industry standard):
\begin{itemize}
\item Regarding LGD: Migration modelling is most likely implemented anyway, so treating default as migration is just a matter of configuration.
\item Regarding PD: Tweaking the intra-sector correlations is straightforward, and the inter-sector correlations just have to be rescaled. Note that uncertainty in migration probabilities will be simultaneously addressed.
\item Regarding correlations: The multivariate normal distribution of the systematic risk factors has to be replaced by a multivariate $t$-distribu\-tion. This might require additional implementation effort but will pay off (eg, will help in fulfilling regulatory requirements as in cf. \cite{EGIM} par. 153). For example, it might be convenient to obtain high correlation between $t$-distributed variables by simultaneous simulation (i.e., summation of the same squared standard normals) of the underlying $\chi^2$-distributions.
\end{itemize}
Of course, in this general setting there are no analytical formulas for Value-at-Risk but the loss distribution has to be approximated by Monte Carlo simulation. Note also that in practice there might be some diversification of model risk (eg, if credit migrations can lead to gains, if there are short positions in the portfolio, or if there are negative inter-sector correlations).

The methods proposed can serve as a starting point for model risk quantification, discussion of effects, and model risk management. Application in practice can lead to interesting effects and conclusions, depending on the specific portfolio and model. For example, idiosyncratic effects might dominate the credit risk computations so that uncertainty in the correlations is a minor issue. Moreover, the type of distributions will play a role: skewed distributions (e.g., around small PDs) will have greater effect than symmetric ones (e.g., around correlations). We hope that readers of this short note will explore in this direction, and thus advance management of model risk for credit portfolio models in the financial industry.


\end{document}